\documentclass[fleqn,10pt]{wlscirep} \title{Initial rigid response and softening transition of highly stretchable kirigami sheet materials} \author[]{Midori Isobe} \author[1,*]{Ko Okumura} \affil[]{Department of Physics and Soft Matter Center, Ochanomizu University, 2--1--1, Otsuka, Bunkyo-ku, Tokyo 112-8610, Japan} \affil[*]{corresponding author: okumura@phys.ocha.ac.jp} 

\begin{abstract}
We study, experimentally and theoretically, the mechanical response of sheet
materials on which line cracks or cuts are arranged in a simple pattern. Such
sheet materials, often called kirigami (the Japanese words, kiri and gami,
stand for cut and paper, respectively), demonstrate a unique mechanical
response promising for various engineering applications such as stretchable
batteries: kirigami sheets possess a mechanical regime in which sheets are
highly stretchable and very soft compared with the original sheets without
line cracks, by virtue of out-of-plane deformation. However, this regime
starts after a transition from an initial stiff regime governed by in-plane
deformation. In other words, the softness of the kirigami structure emerges as
a result of a transition from the two-dimensional to three-dimensional
deformation, i.e., from stretching to bending. We clarify the physical origins
of the transition and mechanical regimes, which are revealed to be governed by
simple scaling laws. The results could be useful for controlling and designing
the mechanical response of sheet materials including cell sheets for medical
regeneration and relevant to the development of materials with tunable
stiffness and mechanical force sensors.

\end{abstract}
\begin{document}

\flushbottom\maketitle
\thispagestyle{empty}

\section*{Introduction}

\begin{figure}[ptb]
\begin{center}
\includegraphics[width=\linewidth]{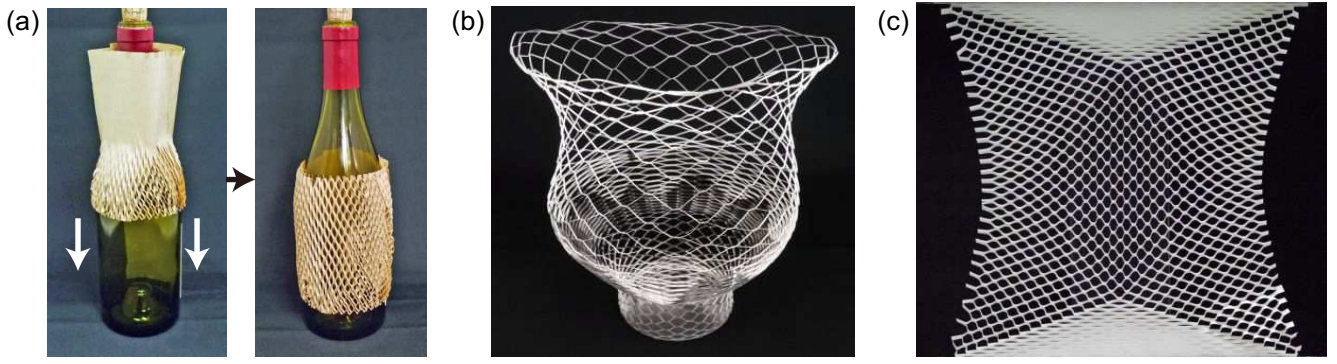}
\end{center}
\caption{(a) A sheet of paper perforated with many cuts, \textit{kirigami},
used for the protection of a bottle of wine. (b) \textit{"Airvase"} (Trafu
Architects, Japan) sold in museum shops worldwide, made from a sheet of
kirigami. (c) Planer stretching of a sheet of kirigami with similar
perforation geometry. The lack of circular symmetry leads to inhomogeneously
stretched cuts.}%
\label{Fig0}%
\end{figure}

Sheet materials, such as paper, plastic film, and metal foil, are a familiar
form of materials and useful in daily life, for example, for wrapping.
However, their unique modes of mechanical responses are highly nontrivial and
have actively been studied mainly from fundamental points of view, which
includes crumpling of paper \cite{Witten1995Science,witten2007stress},
pleating of paper (Miura-ori or Origami)
\cite{mahadevan2005Science,Mahadevan2013PRL,vanHeckOrigami2015PRL}, creasing
of elastomer films \cite{Hayward2012PRL}, wrinkling of thin sheets
\cite{WrinklingScience2007,WrinklingCurtains2011PRL}, and twisting of ribbons
\cite{TwistedRibbon2013PRL}. Quite recently, it has been shown that such
peculiar mechanical responses of sheet materials are also promising for
engineering applications, such as foldable actuators \cite{foldingPNAS2010},
self-folding shape-memory composites \cite{SelfFolding2014Science},
stretchable lithium-ion batteries \cite{Kirigami2015SciRep}, stretchable
electrodes \cite{Kirigami2015NatMat}, stretchable graphens
\cite{GraphenKirigami2015Nature,GraphenKirigami2014PRB}, and integrated solar
tracking \cite{KirigamiSolarNC2015}. One of the key factors in these quite
recent engineering applications is the introduction of many cuts into sheet
materials, often called the kirigami approach in recent papers. In principle,
this approach allows us to design and control the elastic properties of sheet
materials in a highly flexible manner. In fact, supermarkets in Japan often
distribute one who buys bottles of wine with sheets of paper perforated with
regularly arranged cuts to protect the bottles (see Fig. \ref{Fig0}(a)).
Similarly, a Japanese design factory produces "airvase" (Fig. \ref{Fig0}(b))
sold in museum shops worldwide. Figure \ref{Fig0}(c) demonstrates a paper with
similar cuts in planer tension.

However, any simple relations between the mechanical response and arrangements
of cuts have not been explored, although such a relation, if available, could
be useful for designing commercial, engineering, or artistic applications.
Here, we performed a systematic study on the force-extension relation for
sheets of papers with regularly arranged cuts. As a result, we find a number
of regimes for the mechanical response and clarify the physics of the
transition between the first rigid and second soft regimes, at the level of
scaling laws.

\section*{Experiment}

\begin{figure}[ptb]
\begin{center}
\includegraphics[width=0.8\linewidth]{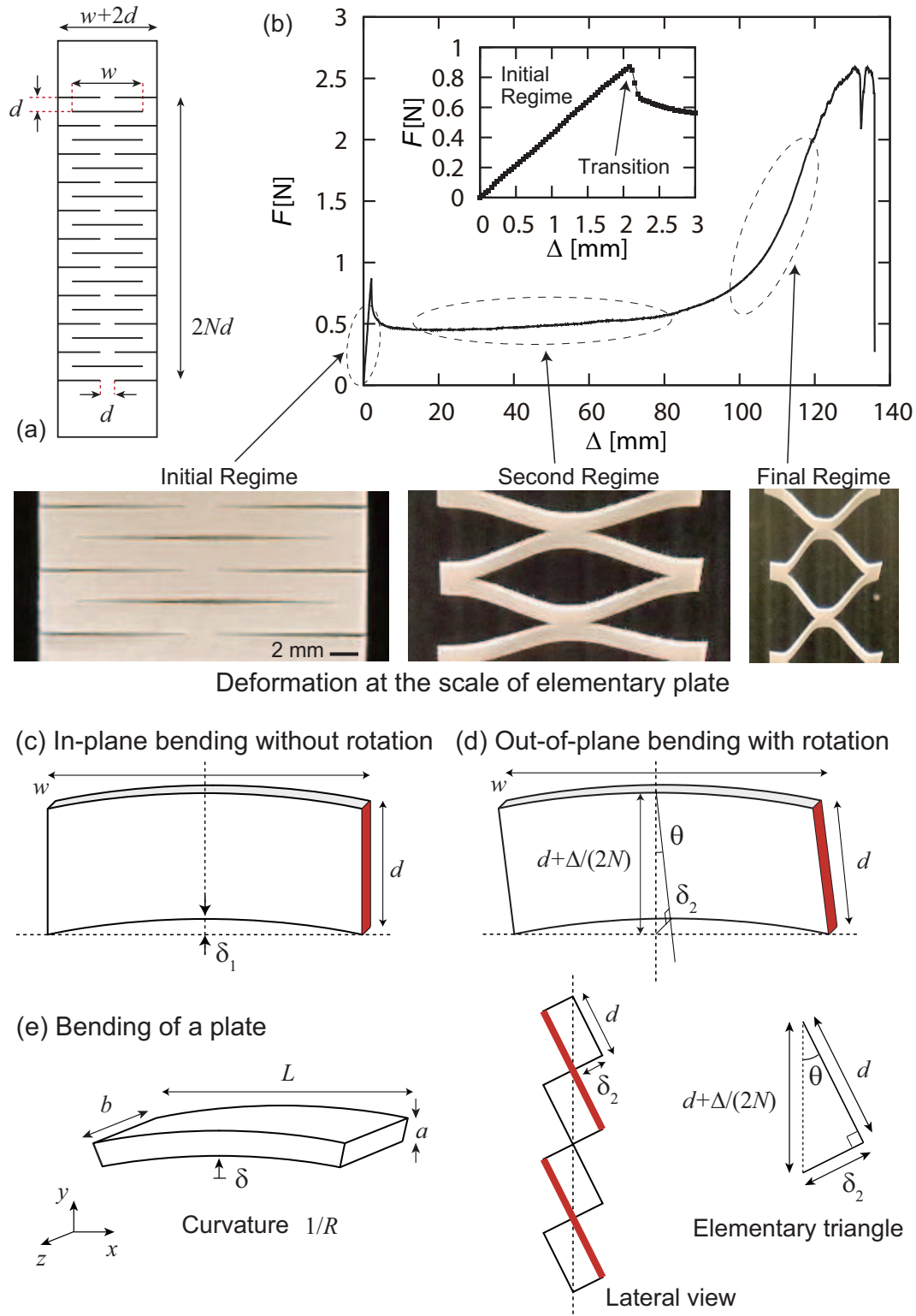}
\end{center}
\caption{(a) Kirigami pattern investigated in the present study. (b) Force $F$
vs. extension $\Delta$. The initial regime shown in the inset is linear, which
is followed by the second soft regime and the final hardening regime. (c)
In-plane deformation of the unit plate in the initial regime. (d) Out-of-plane
deformation in the second regime: perspective view in the top and lateral view
in the left bottom. (e) Illustration of bending of a plate to discuss the
deformation energy.}%
\label{Fig1}%
\end{figure}

In this study, we focus on simple perforation patterns as shown in Fig.
\ref{Fig1}(a) and in the Supporting Information video file. Patterns are
fabricated by a commercial cutting plotter (silhouette CAMEO, Graphtec Corp.).
The patterns are characterized by the length $w$ of each cut ($w\simeq10-30$
mm), the horizontal and vertical spacing $d$ between the cuts ($d\simeq1-5$
mm), with $w$ and $d$ satisfying the condition $w>d$. The number $N$ of cuts
of length $w$ is fixed to $10$, making the sample height to be $2Nd$ as
indicated in Fig. \ref{Fig1}(a). The sample can be regarded as a serial
connection of $2N$ elementary plates characterized by the lengths, $h$, $d$,
and $w+2d\simeq w$. The material of sheet samples is Kent paper (high quality
paper with fine texture mainly used for drafting) of thickness $h$
($h\simeq0.2-0.3$ mm), whose Young's modulus $E$ is measured to be in the
range $E\simeq2.45-3.27$ GPa (with the standard deviations less than $\simeq$
5 \%). We measured force as a function of extension by a force gauge (FGP-0.2,
NIDEC-Shimpo) mounted on an automatic slider system (EZSM6D040, Oriental
Motor) as in the previous studies on fracture
\cite{Kashima2014,SoneMoriJSPS,Shiina06}. The extension speed is fixed to a
slow speed (0.5 mm/s) to remove dynamic effects. In order to minimize
experimental errors, the data in Figs. \ref{Fig3} and \ref{Fig4} below are
obtained within a short period in which temperature and humidity are
relatively stable.

An example of the overall mechanical response is shown in Fig. \ref{Fig1}(b)
with snapshots. In the first regime, which is linear as shown in the inset,
the deformation is restricted in-plane deformation: small stretch occurs as a
result of in-plane bending of elementary plates (a simplified view is shown in
Fig. \ref{Fig1}(c)). In the second regime, out-of-plane bending of elementary
plates accompanied by their rotation of angle $\theta$ allows more stretch
(see Fig. \ref{Fig1}(d) for a simplified view). In the third regime, the
deformation is rather localized near the tips of the cuts, leading to
hardening of the mechanical response and finally to fracture.

\section*{Theory}

The principle results of this paper can be summarized as follows. In the
initial to second regime, the in-plane deformation energy competes with the
out-of-plane deformation energy. The transition between the two regimes occurs
when the two energies become equal. This condition is found to be given by the
following critical extension $\Delta_{c}=2N\delta_{c}$ or critical strain
$\varepsilon_{c}=\delta_{c}/d$:%
\begin{equation}
\delta_{c}\simeq h^{2}/d. \label{e0}%
\end{equation}
In the initial regime ($\Delta<\Delta_{c}$), the response can be described by
the force-extension law or the stress-strain relation, which is linear:%
\begin{equation}
F=K_{1}\Delta\text{ or }\sigma=E_{1}\varepsilon, \label{e1}%
\end{equation}
where $K_{i}=k_{i}/(2N)$ and $E_{i}=k_{i}d/(hw)$ with%
\begin{equation}
k_{1}\simeq Ed^{3}h/w^{3}. \label{e2}%
\end{equation}
In the second regime ($\Delta>\Delta_{c}$), the response becomes quasi-linear:%
\begin{equation}
F=cK_{1}\Delta_{c}+K_{2}\Delta\text{ or }\sigma=cE_{1}\varepsilon_{c}%
+E_{2}\varepsilon\label{e3}%
\end{equation}
with $0<c$ $<1$ where
\begin{equation}
k_{2}\simeq Eh^{3}d/w^{3}. \label{e4}%
\end{equation}
This transition is certainly from hard to soft regime, as confirmed by the
ratio $K_{2}/K_{1}\simeq(h/d)^{2}\ll1$. The observed drop of force at the
transition can be estimated as $(1-c)K_{1}\Delta_{c}$ because $K_{1}\Delta
_{c}\gg K_{2}\Delta$ for small $\theta$. In the following, these relations are
theoretically explained with theoretical limitations and the agreement between
theory and experiment is shown.

In order to understand the mechanical response, we remind the bending energy
of a plate of length $L$, width $b$, and thickness $a$ for the small bending
deflection $\delta$ ($\delta\ll L$) \cite{Landau}:%
\begin{equation}
U(\delta)_{L,a,b}\simeq Ea^{3}b\delta^{2}/L^{3}.\label{eq10}%
\end{equation}
Dimensionally, this is given as follows (see Fig. \ref{Fig1}(e)). We set the
$x$, $y$, and $z$ axes in the direction of $L$, $a$, and $b$, respectively.
The energy per unit volume for a bending of the plate (Young's modulus $E$)
characterized by the curvature $R$ scales as $E\varepsilon^{2}/2$ (this is
exact when Poisson's ratio $\nu$ is zero), where the strain is estimated by
$\varepsilon=((R+y)\varphi-R\varphi)/R\varphi=y/R$ with $\varphi$ the central
angle of the arc in Fig. \ref{Fig1}(e) when the plate occupies the region
$-a/2\leqq y\leqq a/2$. For the deflection $\delta$ of the plate in the $y$
direction, the total bending energy is given by $Lb\int_{-a/2}^{a/2}%
dyE(y/R)^{2}/2$, which leads to Eq. (\ref{eq10}) with the numerical
coefficient 8/3 (i.e., $8/(3(1-\nu^{2}))$ at $\nu=0$ \cite{Landau}), because
the radius of curvature $1/R$ is given by $2\delta/(L/2)^{2}$ in the limit
$\delta\ll L$.

To characterize the mechanical response in the initial regime, we simply
consider superposition of the in-plane bending illustrated in Fig.
\ref{Fig1}(c). By identifying the parameter set $(L,a,b)$ with the set
$(w,d,h)$, we obtain the deformation energy in the initial regime:
\begin{equation}
U_{1}(\Delta)=2NU(\delta_{1})_{w,d,h}\text{ for }\delta_{1},d\ll w
\label{eq11}%
\end{equation}
with $\Delta=2N\delta_{1}$ because our test samples can be regarded as a
serial connection of $2N$ elementary plates. This energy scaling with
$\Delta^{2}$ results in the linear force-extension relation in Eq. (\ref{e1})
with Eq. (\ref{e2}). Here, the stress $\sigma$ and the corresponding elastic
modulus $E_{1}$ are introduced by the definitions $\sigma=F/(hw)$ and
$\sigma=E_{1}\Delta/(2Nd)$.

The mechanical response in the second regime can be estimated by simply
considering superposition of the out-of-plane bending with rotation
illustrated in Fig. \ref{Fig1}(e). With the replacement $(L,a,b)\rightarrow
(w,h,d)$, the deformation energy is given by%
\begin{equation}
U_{2}(\Delta)=2NU(\delta_{2})_{w,h,d}\text{ for }\delta_{2},d\ll w \label{q13}%
\end{equation}
with the relation $\delta_{2}^{2}=(\Delta/2N+d)^{2}-d^{2}$ (see the triangle
in Fig. \ref{Fig1}(d)). The energy in Eq. (\ref{q13}) scaling with
$(\Delta/2N+d)^{2}-d^{2}$ leads to the quasi-linear force proportional to
$\Delta/2N+d$ in Eq. (\ref{e3}) with Eq. (\ref{e4}).

The crossover from the initial to the second regime occurs when the two
energies $U_{1}(\Delta)$ and $U_{2}(\Delta)$ coincide with each other, which
leads to Eq. (\ref{e0}). For a given $\Delta$, the deformation with the
smaller energy is favored, confirming the crossover from $U_{1}(\Delta)$ to
$U_{2}(\Delta)$ at $\Delta=\Delta_{c}$ as $\Delta$ increases. We can show that
the numerical coefficient for Eq. (\ref{e0}) and Eq. (\ref{e2}) are of the
order of unity and $0<c$ $<1$ in Eq. (\ref{e3}), as announced, in a naive
assumption in which the numerical coefficients for Eq. (\ref{eq11}) and Eq.
(\ref{q13}) are both given by $8/(3(1-\nu^{2}))$.

\section*{Experiment and theory}

\begin{figure}[ptb]
\begin{center}
\includegraphics[width=\linewidth]{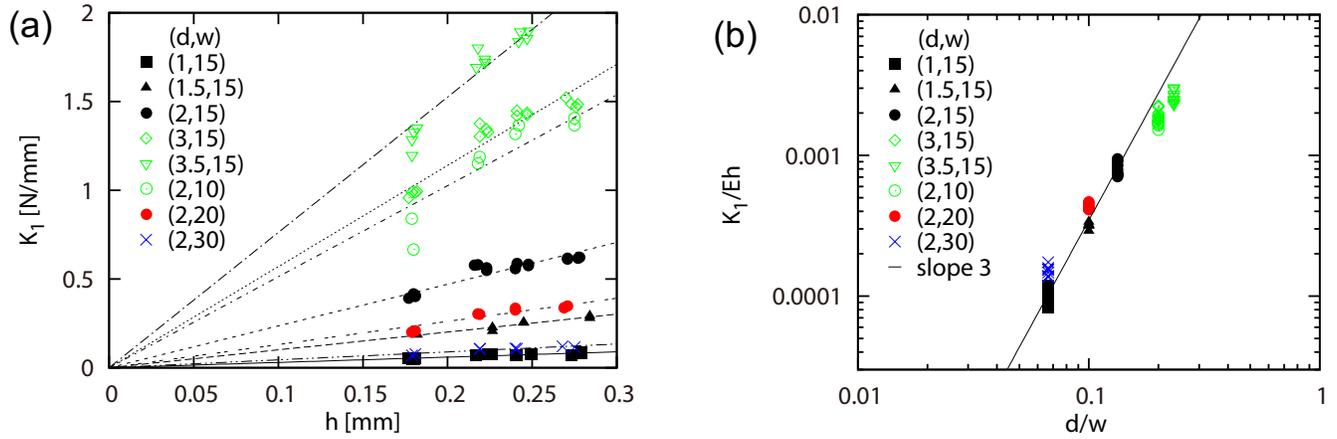}
\end{center}
\caption{(a) Stiffness constant $K_{1}$ vs thickness $h$ for various cut
length $w$ and spacing $d$. The lines are guide for the eyes. (b) $K_{1}/(hE)$
vs $(d/w)^{3}$ demonstrating collapse of the data in (a) by rescaling of the
both axes in (a) according to Eq. (\ref{e2}). The slight deviation of the open
(green) symbols is consistent with the prediction: for these data the
condition $w\leqq5d$ is satisfied whereas the theory requires the condition
$d\ll w$.}%
\label{Fig3}%
\end{figure}

Equation (\ref{e2}) for the stiffness constant $K_{1}$ can be well confirmed
as shown in Fig. \ref{Fig3}(a) and (b). This quantity, experimentally
determined from the slope of a plot as shown in the inset of Fig.
\ref{Fig1}(b), is given as a function of $h$ for various $d$ and $w$ in Fig.
\ref{Fig3}(a). When the two axes are rescaled according to Eq. (\ref{e2}),
namely, $K_{1}/(hE)\simeq(d/w)^{3}$, all the data in Fig. \ref{Fig3}(a)
collapse onto a master curve as shown in Fig. \ref{Fig3}(b), confirming the
theory. The slight discrepancy that can be recognized for the data $w\leqq5d$
is consistent with the prediction because the theory requires the condition
$w\gg d$. This collapse predicts the numerical coefficient for this scaling
law to be $0.346\pm0.006$ (based on the data with $w>5d$), which is of the
order of unity, as expected.

\begin{figure}[ptb]
\begin{center}
\includegraphics[width=\linewidth]{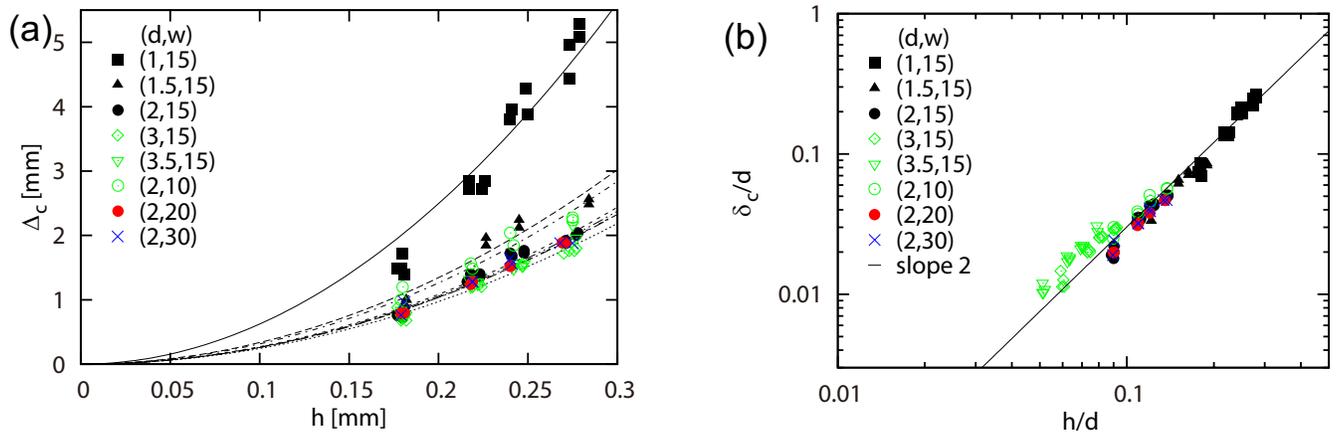}
\end{center}
\caption{(a) Critical spacing $\Delta_{c}$ vs thickness $h$ for various length
$w$ and spacing $d$. The curves are guide for the eyes. (b) $\delta_{c}/d$ vs.
$(h/d)^{2}$ demonstrating collapse of the data in (a) by rescaling of the both
axes in (a) according to Eq. (\ref{e0}). The slight deviation of the open
(green) symbols is again consistent with the prediction.}%
\label{Fig4}%
\end{figure}

Equation (\ref{e0}) for $\delta_{c}$ can also be well confirmed as shown in
Fig. \ref{Fig4}(a) and (b). The critical extension $\Delta_{c}$ can be
estimated as the end point of the initial linear regime as shown in the inset
of Fig. \ref{Fig1}(b). This quantity is given as a function of $h$ for various
$d$ and $w$ in Fig. \ref{Fig4}(a). When the two axes are rescaled according to
Eq. (\ref{e0}), namely, $\delta_{c}/d\simeq(h/d)^{2}$, all the data in Fig.
\ref{Fig4}(a) collapse onto a master curve as shown in Fig. \ref{Fig4}(b),
confirming the theory. The slight discrepancy recognized for the data with
$w\leqq5d$ is again consistent with the prediction. According to this collapse
(of the data with $w>5d$), the numerical coefficient for the scaling law in
Eq. (\ref{e0}) is obtained as $3.02\pm0.05$. This value is of the order of
unity as expected.

\section*{Discussion}

The assumption employed in Eqs. (\ref{eq11}) and (\ref{q13}) that all the
elementary plates behave in the same way may be reasonable at the level of
scaling laws, as strongly supported by the good agreement between theory and
experiment. The extension from this level of description should be examined
further in a separate study.

Although no previous studies are available that focus on the initial rigid
regime and the transition of this regime to the following softer regime, at
the time of writing this paper we find a number of recent related studies
mainly in the engineering community as mentioned the introductory paragraph
(In fact, we started the present study, inspired by our previous study
\cite{AoyanagiOkumura2010} and examples in Fig. \ref{Fig0}). Observations in
the previous studies are qualitatively explicable by our simple theory : (1)
The observations and finite-element-modeling (FEM)\ calculations in the
previous study \cite{Kirigami2015NatMat} focusing on the second regime are
consistent with Eq. (\ref{e4}), confirming that the soft spring constant
increases as $d$ increases and as $w$ decreases. (2) In molecular-dynamics
(MD) simulations performed for a graphen kirigami in the previous study
\cite{GraphenKirigami2014PRB}, the initial rigid regime is practically not
observed, which is consistent with our prediction of the disappearance of the
initial regime in the limit $h\ll d$ (see below).

Our results provide guiding principles in the form of simple scaling laws to
control and design similar kirigami structures. The control and design are
important from two opposite aspects. (1) One aspect is to realize stretchable
sheet materials from stiff materials. In such a case, this initial regime is
unfavorable and Eq. (\ref{e0}) gives a clear principle to reduce the range of
this initial regime: this regime disappears in the limit $h\ll d$. This
prediction is consistent with the previous study \cite{GraphenKirigami2014PRB}
as mentioned above. (2) The opposite aspect is positive utilization of the
initial rigid regime, for which we propose two new directions of applications
(Note, however, that even in the "rigid regime" the kirigami sheet is already
significantly soft compared with the original material as seen from the factor
$(d/w)^{3}$ in Eq. (\ref{e2})). One possibility is fine tuning of the elastic
constant of sheet materials. By virtue of Eq. (\ref{e2}), we could design the
elasticity of sheet materials at will (in the initial regime). It would be
interesting, for example, to use relatively thick sheet materials, widening
the range of the initial regime. Another possibility is the application for
mechanical force sensors. Because of the sudden elongation at the critical
length and force, we could design force sensors on the basis of Eq. (\ref{e0})
with Eq. (\ref{e1}). In addition, Eq. (\ref{e4}) for small deformation should
be useful to design and control the soft response of the kirigami sheets (note
that Eq. (\ref{e4}) qualitatively justify observations in the previous study
\cite{Kirigami2015NatMat} as mentioned above). A promising example of
applications relevant to the present study would be the use of the kirigami
structure for cell sheets, which have received considerable attention in
regenerative medicine \cite{CellSheet2007}.

As pointed out already in the above, our prediction is quite consistent with
the FEM results in the previous study \cite{Kirigami2015NatMat}, whereas the
FEM approach and the model proposed here have advantages and disadvantages.
The FEM approach predicts the results numerically, whereas the present model
predicts the results analytically but without a precise prediction for a
numerical coefficient. The numerical coefficient is precisely determined only
through a comparison with experimental data. But once this is done, the
present model predicts the results numerically in a wide range of important
physical parameters without any technical efforts required for the FEM
approach. In addition, the present approach provides physical insights into
the phenomenon in a clearer manner, giving simple guiding principles for
designing the kirigami structure. However, the present prediction can be used
only for the cases satisfying the required conditions, such as $d\ll w$,
unlike the FEM analysis; the distribution of strain and stress is only
available in the FEM analysis.

We consider that the nonlinearity in the stress-strain relationship of sheet
materials may not strongly affect the physical pictures provided in the
present study. Most of materials certainly possess such nonlinearity for large
stresses, whereas the kirigami structure contains many cuts at the tips of
which stress could be high. Thus, it would be natural to ask how such
nonlinear effects affect the present framework. This problem could be in
general nontrivial. However, we expect that nonlinear effects tend to be
suppressed in the kirigami structure at least in the initial and second
regime, which are the focus of the present study. This is because of the
following reasons: (1) In the initial regime the deformation is generally very
small. (2) In the second regime the apparent deformation is large but the
basic mode of deformation is still bending, whereas bending is intrinsically
related to small deformation especially when the plate is thin. This
expectation for a non-significant role of the nonlinearity is supported by the
agreement between theory and experiment in the present study. However, we
experimentally observed that the linearity of the slope in the initial regime
could be slightly deteriorated near the transition point in certain cases
(although such data points can be well explained by our theory). This might be
a slight effect of the nonlinearity in the stress-strain relationship of the
sheet material.

\section*{Conclusion}

We investigated the mechanical response of a simple and representative
kirigami structure that remarkably changes original mechanical properties in a
systematic way. As a result, we found simple scaling laws that govern the
stiffness of the initial regime and the consecutive softening transition. This
transition was revealed to be a transition from the two-dimensional to
three-dimensional deformation, i.e., from stretching to bending. The result
obtained here could be useful as design principles for simple kirigami
structures. Upon seeing the recent surge of engineering utilization of
kirigami structures, we envision that the present results would be useful for
various applications, as well as for fundamental understanding of the
mechanics of sheet materials.

\bibliographystyle{naturemagMy}
\bibliography{C:/Users/okumura/Documents/main/papersPDF/granular,C:/Users/okumura/Documents/main/papersPDF/fracture,C:/Users/okumura/Documents/main/papersPDF/wetting}

\section*{Acknowledgements}

We thank Atsushi Takei (Ochanomizu University) for discussions and useful
comments. This research was partly supported by Grant-in-Aid for Scientific
Research (A) (No. 24244066) of JSPS, Japan, and by ImPACT Program of Council
for Science, Technology and Innovation (Cabinet Office, Government of Japan).

\section*{Author contributions statement}

K.O. and M.I. conceived the experiments, M.I. conducted the experiments, M.I
and K.O. analyzed the results, M.I. and K.O. prepared the figures and graphs,
K.O wrote the manuscript. All authors reviewed the manuscript.

\section*{Additional information}

Competing financial interests: The authors declare no competing financial interests.


\end{document}